\documentclass[preprint,12pt]{elsarticle}

\usepackage{amssymb}
\usepackage{graphicx}
\usepackage{dcolumn}
\usepackage{bm}
\usepackage{float,color}
 \usepackage{amsthm}

\journal{Nuclear Physics A}

\begin{document}

\begin{frontmatter}

\title{Covariant description of shape evolution and shape coexistence in neutron-rich nuclei at $N\approx60$}
\author{J. Xiang, Z. P. Li, Z. X. Li}
\address{School of Physical Science and Technology, Southwest
University, Chongqing 400715, China}
\author{J. M. Yao}
\address{School of Physical Science and Technology, Southwest University, Chongqing, 400715 China}
\address{Physique Nucl\'eaire Th\'eorique,
             Universit\'e Libre de Bruxelles, C.P. 229, B-1050 Bruxelles,
             Belgium}
 \author{J. Meng}
\address{State Key Laboratory of Nuclear Physics and Technology,
  School of Physics, Peking University, Beijing 100871, China}
\address{School of Physics and Nuclear Energy Engineering, Beihang University, Beijing 100191, China}

\begin{abstract}
 The shape evolution and shape coexistence phenomena in neutron-rich nuclei at $N\approx60$,
 including Kr, Sr, Zr, and Mo isotopes, are studied in the covariant density functional theory (DFT) with
 the new parameter set PC-PK1. Pairing correlations are treated using the BCS approximation with a separable pairing
force. Sharp rising in the charge radii of Sr and Zr isotopes at
$N=60$ is observed and shown to be related to the rapid changing in
nuclear shapes. The shape evolution is moderate in neighboring Kr
and Mo isotopes. Similar as the results of previous
Hartree-Fock-Bogogliubov (HFB) calculations with the Gogny force,
triaxiality is observed in Mo isotopes and shown to be essential to
reproduce quantitatively the corresponding charge radii. In
addition, the coexistence of prolate and oblate shapes is found in
both $^{98}$Sr and $^{100}$Zr. The observed oblate
 and prolate minima are related to the low single-particle energy level
 density around the Fermi surfaces of neutron and proton
 respectively. Furthermore, the 5-dimensional (5D) collective Hamiltonian determined by the
 calculations of the PC-PK1 energy functional is solved for $^{98}$Sr and $^{100}$Zr. The resultant excitation
 energy of $0^+_2$ state and E0 transition strength $\rho^2(E0;0^+_2\rightarrow0^+_1)$
 are in rather good agreement with the data. It is found that the lower barrier
 height separating the two competing minima along the $\gamma$ deformation
 in $^{100}$Zr gives rise to the larger $\rho^2(E0;0^+_2\rightarrow0^+_1)$
 than that in $^{98}$Sr.
\end{abstract}


\begin{keyword}

 Covariant density functional \sep shape evolution and shape
 coexistence \sep charge radii \sep neutron-rich Kr, Sr, Zr, Mo isotopes

\end{keyword}

\end{frontmatter}
 \section{\label{sec:level2}Introduction}

In recent decades, the evolution of nuclear shapes along isotopic
and isotonic chains in neutron-rich nuclei at $N\approx60$ has
attracted many attentions. The sudden onset of quadrupole
deformation in neutron-rich Sr and Zr isotopes at the neutron number
$N=60$ is of particular interest. Such a rapid shape evolution has
been deduced from the abrupt changing of lifetimes of $2^+_1$
states~\cite{Mach91,Goodin07} as well as the quadrupole moments of
rotational bands~\cite{Urban01}. Besides, the excitation energies of
$2^+_1$ states~\cite{NNDC}, two-neutron separation
energies~\cite{Hager06}, and mean-square charge
radii~\cite{Charlwood09} exhibit a dramatic change between $N=58$
and 60 in Sr and Zr isotopes. Very recently, the systematic of the
$2^+_1$ states in Kr isotopes has been extended up to
$N=60$~\cite{Marginean09}, at which nucleus, the energy of the first
excited state drops down suddenly by $\sim400$~keV. It indicates
that the shape transition is also rather abrupt in Kr isotopes.
However, the measured charge radii was shown to be increasing
moderately with the neutron number at $N=60$~\cite{Keim95}.

Essential to the understanding of this dramatic shape evolution is
the coexistence of different shapes in the two lowest $0^+$ states.
Shape coexistence phenomena at low energy in Sr and Zr isotopes
around $N=60$ have been shown in many experimental measurements. In
Ref.~\cite{Jung80}, Jung \emph{et al.} discovered two low-lying
$0^+$ states in $^{96}$Sr at 1229 and 1465~keV respectively.  Later
on, an extremely strong electric monopole transition of
$\rho^2(E0)=0.18$ was observed between the first two $0^+$ states
~\cite{Kawade82,Lhersonneau94}. The analysis of $B(E2)$ and
$\rho^2(E0)$ values for both $^{98}$Sr and $^{100}$Zr by Mach
\emph{et al}. indicates that these two nuclei have very similar
structures~\cite{Mach89}. Schussler~\emph{et al.} discovered a very
low-lying $0^+$ state at $215.5$ keV in $^{98}$Sr. The transition
probabilities, the reduced E0 matrix element and the observed level
structure suggest the coexistence of a quadrupole deformed ground
state and a spherical excited $0^+$ state in
$^{98}$Sr~\cite{Schussler80}. To clarify such a picture of shape
coexistence, an experiment to measure the spectroscopic quadrupole
moment of the $2^+_1$ state has been proposed~\cite{Clement10}.

On the theoretical side, the shape evolution around $N=60$ has been
studied extensively with various theoretical models, including the
phenomenological
models~\cite{Federman78,Kumar85,Galeriu86,Michiaki90,Moller95,Skalski97,Xu02,Sonia08},
the interacting boson model~\cite{Gracia05}, the modern shell
model~\cite{Sieja09} and the self-consistent mean-field models with
the Skyrme force~\cite{Bonche85,Skalski93,Bender06}, the Gogny
force~\cite{Delaroche10,CEA,Guzman10} as well as the effective
relativistic Lagrangian~\cite{Lalazissis99}. Most of these models
have shown the increasing of deformations up to $N=60$ and indeed
found the competing prolate and oblate minima. However, the subtle
balance between these two minima depends on the details of
calculations.

In recent years, nuclear covariant DFT has achieved great success in
the description of ground state properties of both spherical and
deformed nuclei all over the nuclear
chart~\cite{Reinhard89,Ring96,Vretenar05,Meng06}. In particular, the
covariant DFT theory with a point-coupling interaction has recently
attracted more and more attention~\cite{Niksic11}. It shows great
advantages in the extension for nuclear low-lying excited states by
using projection techniques~\cite{Yao09}, generator coordinate
methods~\cite{Niksic06,Yao10,Yao11} and collective
Hamiltonian~\cite{Niksic09}. In this framework, there are several
popular parameter sets, including PC-F1~\cite{Burvenich},
DD-PC1~\cite{Niksic}, and PC-PK1~\cite{Zhao10}. Among these
parameter sets, the PC-PK1 was proposed very recently by fitting to
observables of 60 selected spherical nuclei, including the binding
energies, charge radii, and empirical pairing gaps. The success of
PC-PK1 has been illustrated in the description of infinite nuclear
matter and finite nuclei for both ground-state and low-lying excited
states. Furthermore, the PC-PK1 provides a good description for the
isospin dependence of binding energy along either isotopic or
isotonic chain.

Recently, a separable pairing force with two universal parameters
has been introduced, which was adjusted to reproduce the pairing
properties of the Gogny force D1S in nuclear matter~\cite{Tian091}.
The separable pairing force has been shown to be successful in the
description of nuclear matter~\cite{Tian091}, spherical and deformed
nuclei~\cite{Tian092,Tian093,Niksic10,Li10}. Therefore, in this
work, we would like to use the PC-PK1 parameter set together with
the separable force to perform a systematic calculation for the
neutron-rich Kr, Sr, Zr, and Mo isotopes. The shape evolution and
shape coexistence in this region will be examined.

The theoretical framework for the relativistic point-coupling model
with the separable pairing force is described in Sec.~\ref{Sec.I}.
The shape evolution in neutron-rich Kr, Sr, Zr, and Mo isotopes at
$N\approx60$ and shape coexistence phenomena in $^{98}$Sr and
$^{100}$Zr will be discussed in Sec.~\ref{Sec.II}. Finally, a
summary is given in Sec.~\ref{summary}.


\section{The model}
\label{Sec.I}

 In the covariant DFT with point-coupling interaction,
 the energy functional has the following form~\cite{Burvenich,Zhao10},
\begin{eqnarray}
 \label{EDF}
 E_{\rm RMF}
 &=& \sum\limits_k\int d\mathbf{r}~v_k^2\bar\psi_k(\mathbf{r})(-i{\bf \gamma}\cdot{\bf\nabla}+m)\psi_k(\mathbf{r}) \nonumber\\
 & & +\int d\mathbf{r}~{\left(\frac{\alpha_S}{2}\rho_S^2+\frac{\beta_S}{3}\rho_S^3 +
     \frac{\gamma_S}{4}\rho_S^4+\frac{\delta_S}{2}\rho_S\triangle \rho_S \right.}\nonumber\\
 & & +{\left.\frac{\alpha_V}{2}j_\mu j^\mu + \frac{\gamma_V}{4}(j_\mu j^\mu)^2 +
     \frac{\delta_V}{2}j_\mu\triangle j^\mu \right.} \nonumber\\
 & & +\left. \frac{\alpha_{TV}}{2}j^{\mu}_{TV}(j_{TV})_\mu+\frac{\delta_{TV}}{2}
     j^\mu_{TV}\triangle (j_{TV})_{\mu}\right. \nonumber\\
 & & +\left.\frac{1}{4}F_{\mu\nu}F^{\mu\nu}-F^{0\mu}\partial_0A_\mu+e\frac{1-\tau_3}{2}j_\mu A^\mu\right),
\end{eqnarray}
where $e$ is the charge unit for protons and it vanishes for
neutrons. The energy functional (\ref{EDF}) contains 9 coupling
constants $\alpha_S$, $\alpha_V$, $\alpha_{TV}$, $\beta_S$,
$\gamma_S$, $\gamma_V$, $\delta_S$, $\delta_V$ and $\delta_{TV}$.
The subscripts indicate the symmetry of the couplings: $S$ stands
for scalar, $V$ for vector, and $T$ for isovector, while the symbol
refer to the additional distinctions: $\alpha$ refers to
four-fermion term, $\delta$ to derivative couplings, and $\beta$ and
$\gamma$ to the third- and fourth-order terms, respectively.

The local densities and currents in the energy functional
(\ref{EDF}) are determined by,
\begin{eqnarray}
  \label{E13a}
  \rho_S(\mathbf{r})          &=&\sum_{k }v^2_k\bar\psi_k(\mathbf{r})\psi_k(\mathbf{r}),\\
  \label{E13b}
  j^\mu(\mathbf{r})           &=&\sum_{k }v^2_k\bar\psi_k(\mathbf{r})\gamma^\mu\psi_k(\mathbf{r}),\\
  \label{E13c}
  \vec j^\mu_{TV}(\mathbf{r}) &=&\sum_{k }v^2_k\bar\psi_k(\mathbf{r})\vec\tau\gamma^\mu\psi_k(\mathbf{r}).
\end{eqnarray}

Minimizing the energy functional (\ref{EDF}) with respect to
$\bar\psi_k$, one obtains the Dirac equation for the single nucleons
\begin{equation}
 \label{DiracEq}
   [\gamma_\mu(i\partial^\mu-V^\mu)-(m+S)]\psi_k=0.
\end{equation}
The single-particle effective Hamiltonian contains local scalar
$S(\bm{r})$ and vector $V^\mu(\bm{r})$ potentials,
\begin{equation}
\label{potential}
  S(\bm{r})    =\Sigma_S, \quad
  V^\mu(\bm{r})=\Sigma^\mu+\vec\tau\cdot\vec\Sigma^\mu_{TV},
\end{equation}
where the nucleon isoscalar-scalar $\Sigma_S$, isoscalar-vector
$\Sigma^\mu$ and isovector-vector $\vec\Sigma^\mu_{TV}$
self-energies are given in terms of the various densities and
currents,
\begin{eqnarray}
 \Sigma_S           &=&\alpha_S\rho_S+\beta_S\rho^2_S+\gamma_S\rho^3_S+\delta_S\triangle\rho_S,\\
 \Sigma^\mu         &=&\alpha_Vj^\mu_V +\gamma_V (j^\mu_V)^3
                      +\delta_V\triangle j^\mu_V + e A^\mu,\\
 \vec\Sigma^\mu_{TV}&=& \alpha_{TV}\vec j^\mu_{TV}+\delta_{TV}\triangle\vec j^\mu_{TV}.
\end{eqnarray}
For a system with time reversal invariance, the space-like
components of the currents and the vector potential vanish.
Furthermore, one can assume that the nucleon single-particle states
do not mix isospin, that is, the single-particle states are
eigenstates of $\tau_3$. Therefore only the third component of
isovector potentials $\vec\Sigma^\mu_{TV}$ survives. The Coulomb
field $A_0$ is determined by Poisson's equation.

Pairing correlations between nucleons are treated using the BCS
approximation with a pairing force separable in momentum space,
i.e., $\langle k\vert V^{^1S_0}\vert k^\prime\rangle=-G p(k) p(k')$,
introduced in Ref.~\cite{Tian091} with a Gaussian ansatz $p(k) =
e^{-a^2k^2}$. The two parameters $G$ and $a$ have been adjusted to
reproduce the pairing properties of the Gogny force D1S in nuclear
matter. The obtained values for the parameters are $G=-728\;{\rm
MeV\cdot fm}^3$ and $a=0.644\;{\rm fm}$.

In the coordinate space, the separable pairing force takes the
following form,
 \begin{equation}
 \label{pp-force}
 V(\mathbf{r}_1,\mathbf{r}_2,\mathbf{r}_1^\prime,\mathbf{r}_2^\prime)
 =G\delta\left(\mathbf{R}-\mathbf{R}^\prime \right)P(\mathbf{r})P(\mathbf{r}^\prime)
 \frac{1}{2}\left(1-P^\sigma \right),
 \end{equation}
where $\mathbf{R}=\frac{1}{2}\left(\mathbf{r}_1+\mathbf{r}_2\right)$
and $\mathbf{r}=\mathbf{r}_1-\mathbf{r}_2$ are the center-of-mass
and the relative coordinates respectively. $P(\mathbf{r})$ is the
Fourier transform of $p(k)$,
 \begin{equation}
 P(\mathbf{r}) = \frac{1}{\left(4\pi a^2  \right)^{3/2}}e^{-\mathbf{r}^2/4a^2}.
\end{equation}
The pairing force has finite range, and it can preserve
translational invariance due to the presence of the factor
$\delta\left(\mathbf{R}-\mathbf{R}^\prime \right)$. Even though
$\delta\left(\mathbf{R}-\mathbf{R}^\prime \right)$ implies that this
force is not completely separable in coordinate space, the
corresponding antisymmetrized $pp$ matrix elements can be
represented as a sum of a finite number of separable terms in the
basis of a three-dimensional (3D) harmonic oscillator (HO):
 \begin{equation}
 \label{eq:matrix-element}
 \langle\alpha\bar\beta|V|\gamma\bar\delta\rangle =
 G\sum_{N_x=0}^{N_x^0}\sum_{N_y=0}^{N_y^0}\sum_{N_z=0}^{N_z^0}
 (V_{\alpha \bar{\beta}}^{N_xN_yN_z})^*V_{\gamma \bar{\delta}}^{N_xN_yN_z} \; ,
 \end{equation}
where $N_x$, $N_y$, and $N_z$ are the quantum numbers of the
corresponding one-dimensional (1D) HO in the center-of-mass frame.
The summations over $N_x$, $N_y$, and $N_z$ are restricted to finite
terms with cutoffs $N_x^0, N_y^0$, and $N_z^0$ respectively. The
convergence with respect of the cutoffs has to be checked in
calculations. $V_{\alpha \bar{\beta}}^{N_xN_yN_z}$ represents the
single-particle matrix element in the 3D HO basis. In this case, the
pairing field can be written as a sum of a finite number of
separable terms
\begin{equation}\label{pairing gap}
 \Delta_{\alpha \bar\beta}
    =G\sum_{N_x=0}^{N_x^0}\sum_{N_y=0}^{N_y^0}
     \sum_{N_z=0}^{N_z^0}  (V_{\alpha \bar\beta}^{N_xN_yN_z})^\ast  P_{N_xN_yN_z},
\end{equation}
with the coefficients
 \begin{equation}
  P_{N_xN_yN_z}= \sum_{\gamma \delta>0}  V_{\gamma \bar\delta}^{N_xN_yN_z}
     \kappa_{\gamma \bar\delta},
 \end{equation}
where $\kappa_{\gamma \bar\delta}$ is the matrix element of pairing
tensor.  The expression of $V_{\gamma \bar\delta}^{N_xN_yN_z}$ has
been derived in Ref.~\cite{Niksic10}.

In the BCS approximation, the pairing gap $\Delta_k$ for each
single-particle state $\psi_k$ is finally determined as follows,
\begin{equation}
 \Delta_k = \sum_{\alpha\bar\beta}\Delta_{\alpha\bar\beta}F_{k\alpha}F_{k\beta},
\end{equation}
where $F_{k\alpha}$ is the expansion coefficient for the large
component in Dirac spinor $\psi_k$ on the 3D HO basis. The resultant
pairing energy is given by
\begin{equation}
 E_{\rm pair}= G  \sum_{N_x=0}^{N_x^0}\sum_{N_y=0}^{N_y^0}
     \sum_{N_z=0}^{N_z^0}  (P_{N_xN_yN_z})^\ast  P_{N_xN_yN_z}.
\end{equation}

The center-of-mass correction to the energy is considered
microscopically with both the direct and exchange terms,
\begin{equation}
  E_{\rm c.m.}=-\frac{\langle\hat{P}^2_{\rm c.m.}\rangle}{2mA},
\end{equation}
 where $m$ is the mass of nucleons. $A$ is mass number
 and $\hat P_{\rm cm}=\sum_i^A \hat p_i$ is the total momentum in the
 c.m. frame.

The total nuclear energy is determined by
\begin{equation}
E_{\rm tot}=E_{\rm RMF}+E_{\rm pair}+E_{\rm c.m.}
\end{equation}
The potential energy surface (PES) in the plane of deformation
variables is obtained by imposing a quadratic constraint on the mass
quadrupole moments
\begin{equation}
\langle
H\rangle+\sum\limits_{\mu=0,2}C_{2\mu}(\langle\hat{Q}_{2\mu}\rangle-q_{2\mu})^2
\end{equation}
where $\langle H\rangle$ is the total energy, and
$\langle\hat{Q}_{2\mu}\rangle$ denotes the expectation value of the
mass quadrupole operator:
\begin{eqnarray}
\hat{Q}_{20}&=&2z^2-x^2-y^2\\
\hat{Q}_{22}&=&x^2-y^2
\end{eqnarray}
Here $q_{2\mu}$ is the constrained value of the quadrupole moments,
and $C_{2\mu}$ the corresponding stiffness constant~\cite{Ring80}.

 \section{Results and discussion}
 \label{Sec.II}
The new parametrization PC-PK1~\cite{Zhao10} and the separable
pairing force~\cite{Tian091} are adopted in the particle-hole
channel and the particle-particle channel respectively. Parity,
$D_{2}$ symmetry, and time-reversal invariance are imposed. The
Dirac equation is solved by expanding in the basis of eigenfunctions
of a 3DHO in Cartesian coordinate with $12$ major shells, which are
found to be sufficient to obtain a reasonably converged mean-field
PES.

\subsection{Shape evolution in neutron-rich Sr isotopes} \label{Sec.II-1}

\begin{figure}[]
\centering
\includegraphics[scale=0.3]{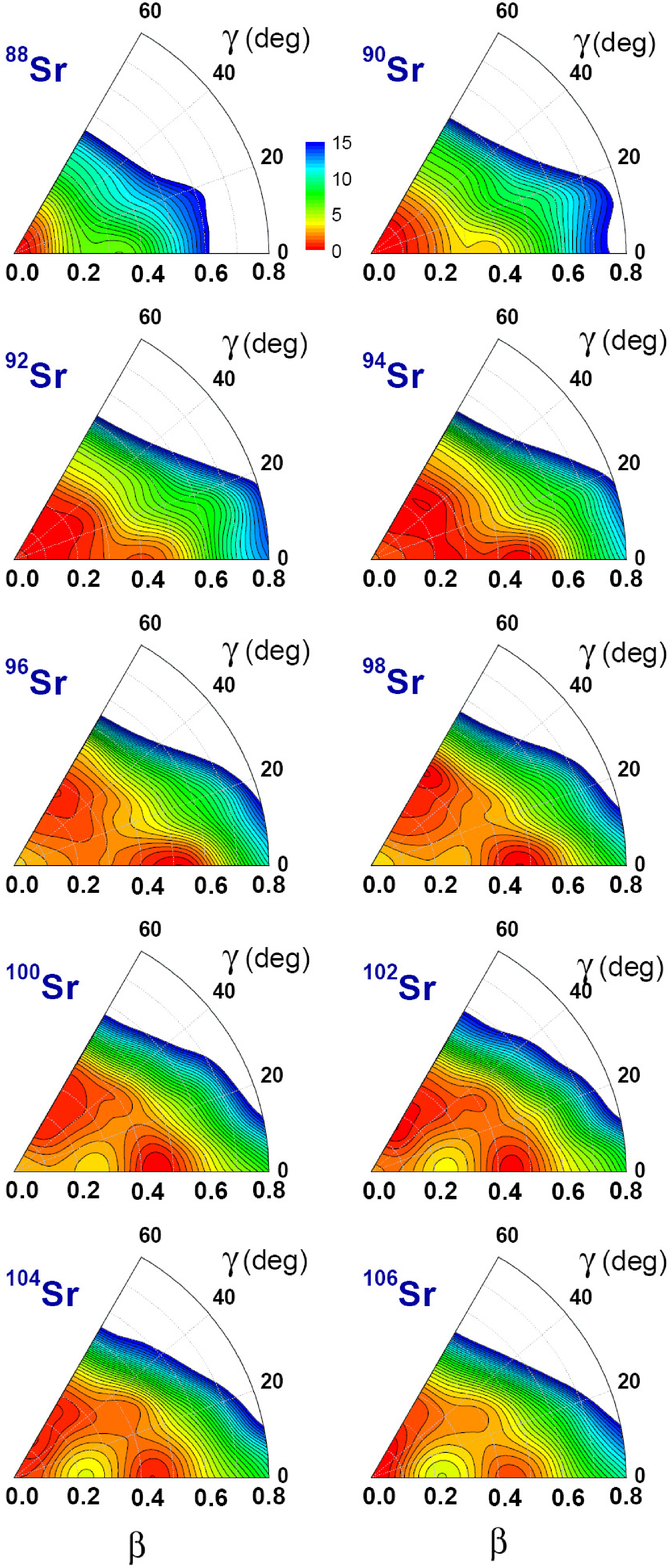}
\caption{\label{fig1}(Color online) The potential energy surfaces of
the even-even $^{88-106}$Sr isotopes in the $\beta$-$\gamma$ plane
from the constrained relativistic mean-field (RMF) plus BCS
calculations. All energies are normalized with respect to the tot
energy of the absolute minimum. The energy difference between
neighboring contour lines is 0.5~MeV.}
\end{figure}

Figure~\ref{fig1} displays the PESs of even-even $^{88-104}$Sr in
$\beta$-$\gamma$ plane, normalized to the total energy of absolute
minimum. The energy difference between neighboring contour lines is
0.5~MeV. The PESs in Fig.~\ref{fig1} show a clear picture for the
evolution of shapes in $^{88-104}$Sr. Starting from a well spherical
shape of $^{88}$Sr, the spherical (global) minima in $^{90,92}$Sr
become soft against the distortion towards oblate shape. In the
mean-time, the prolate (second) minimum comes down. When the neutron
number increases from $N=56$ to $N=60$, the global minimum is
shifted to the oblate side with large deformation. Meanwhile, the
prolate minimum becomes deep and competing with the oblate minimum
at $N=60$. A triaxial barrier with the height $\sim2.22$~MeV
separates these two competing minima in $^{98}$Sr. Beyond the
$N=60$, the structure of the energy maps is stable, that is, a soft
oblate minimum against the distortion towards spherical shape
coexists with a well prolate one.

\begin{figure}[]
\begin{center}
\includegraphics[width=0.4\textwidth]{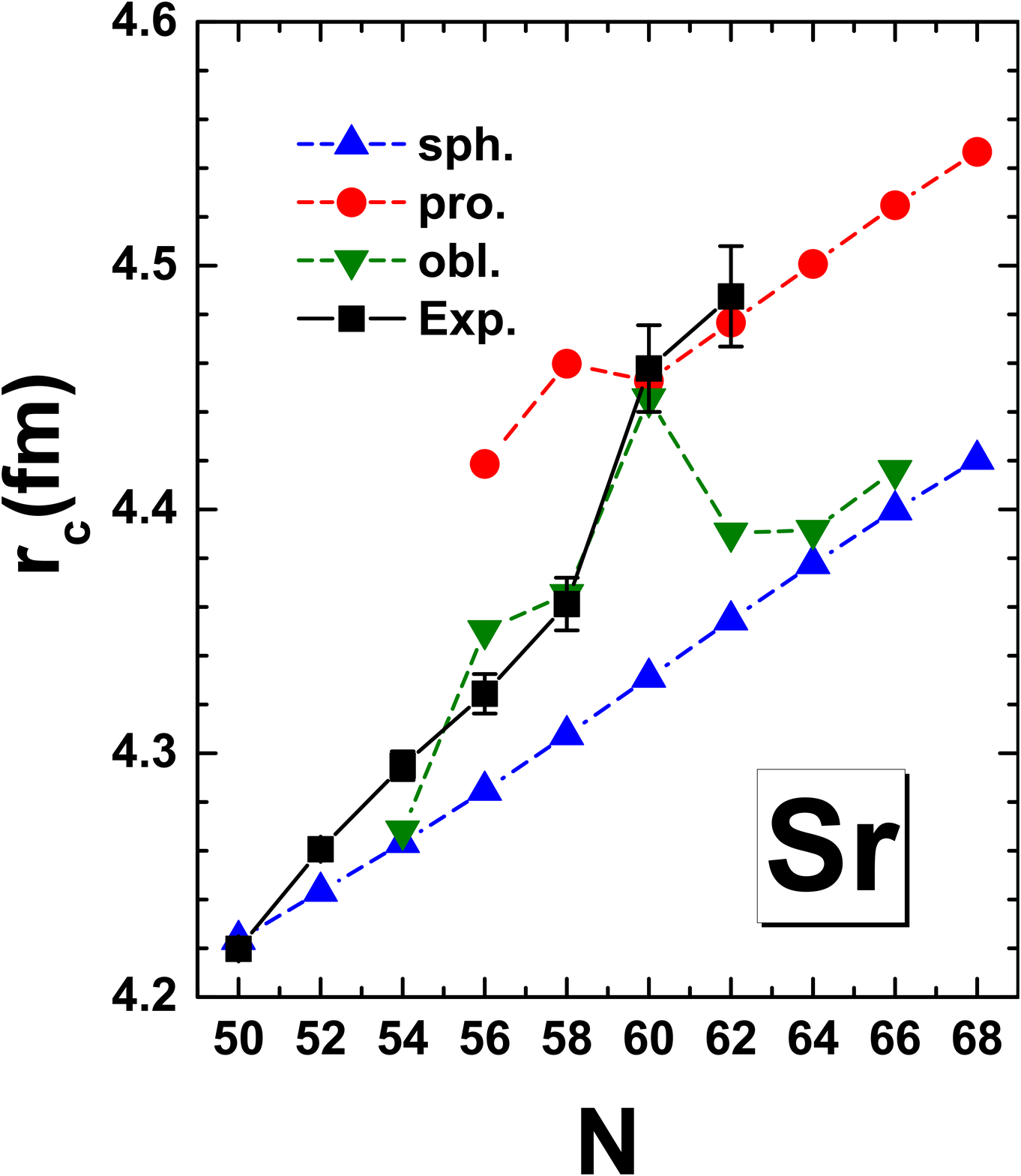}
\end{center}
\caption{\label{fig01}(Color online) The evolution of the nuclear
charge radii in Sr isotopes. The calculated values corresponding to
the spherical (up triangles), prolate (circles), and oblate (down
triangles) local minima in the PESs (Fig. \ref{fig1}) are plotted as
functions of neutron number. The squares with error bars denote the
experimental data \cite{Angeli}.}
\end{figure}

The evolution of the nuclear charge radii in neutron-rich Sr
isotopes can be seen in Fig.~\ref{fig01}, where the calculated
charge radii corresponding to the spherical, prolate, and oblate
local minima in the PESs of even-even $^{88-104}$Sr (c.f.
Fig.~\ref{fig1}) are plotted as functions of neutron number. For
$N\geq54$, the charge radii of spherical shapes are given as well
for comparison. It is shown that the charge radii of the spherical
and prolate shapes increase smoothly in the similar slop with the
neutron number. The difference in the two charge radii is about
$0.12$~fm, originating from the effect of prolate deformation.
Moreover, it means that the deformation of prolate minimum is nearly
the same when the neutron number increases from $N=60$ to $N=68$. On
the contrary, the charge radius corresponding to the oblate minimum
changes rapidly with the neutron number. In particular, a sudden
rising of charge radius from $N=58$ to $N=60$ and a sudden dropping
from $N=60$ to $N=62$ are due to the increasing and decreasing of
the oblate deformation from $\beta=-0.25$ to $\beta=-0.35$ and back
to $\beta=-0.2$. Moreover, it is shown in Fig.~\ref{fig01} that the
charge radii of prolate and oblate minima in $^{98}$Sr are similar
due to the similar size of quadrupole deformation. Comparing with
the available data for charge radii, one can draw a shape evolution
picture for the ground states of even-even $^{88-100}$Sr, namely,
from spherical shape ($^{88}$Sr) to more oblate shape ($^{94}$Sr),
oblate and prolate coexistence ($^{98}$Sr) and finally more prolate
shape ($^{100}$Sr).

\begin{figure}[]
\begin{center}
\includegraphics[scale=0.3]{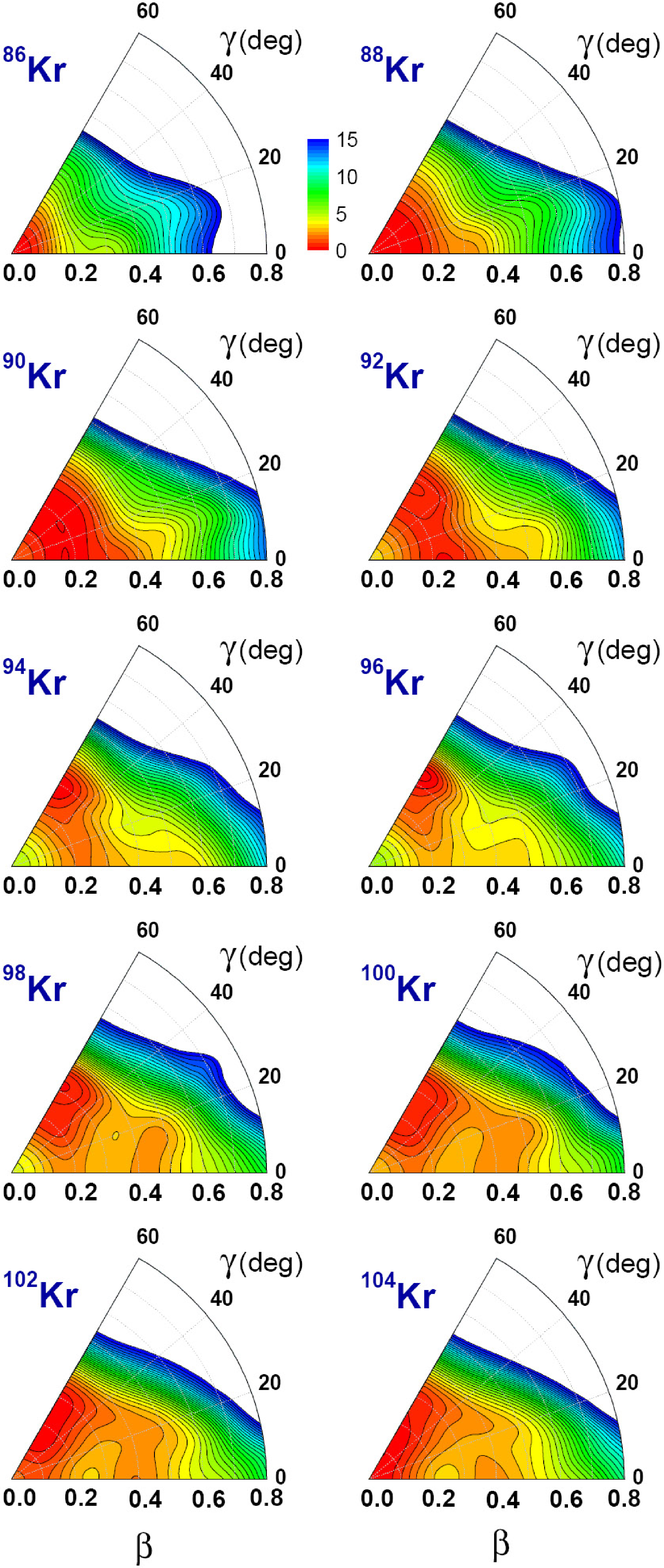}
\end{center}
\caption{\label{fig2}(Color online) Same as the Fig.~\ref{fig1}, but
for the isotopes $^{86-104}$Kr.}
\end{figure}

\begin{figure}[]
\begin{center}
\includegraphics[scale=0.3]{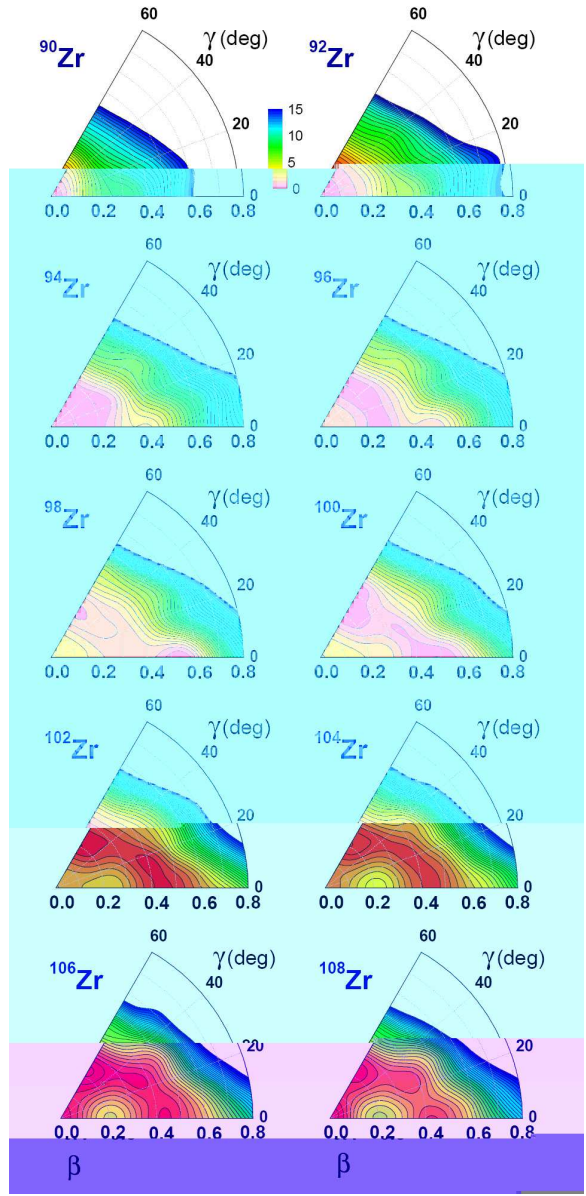}
\end{center}
\caption{\label{fig3}(Color online) Same as the Fig.~\ref{fig1}, but
for the isotopes $^{90-108}$Zr.}
\end{figure}

\begin{figure}[]
\begin{center}
\includegraphics[scale=0.29]{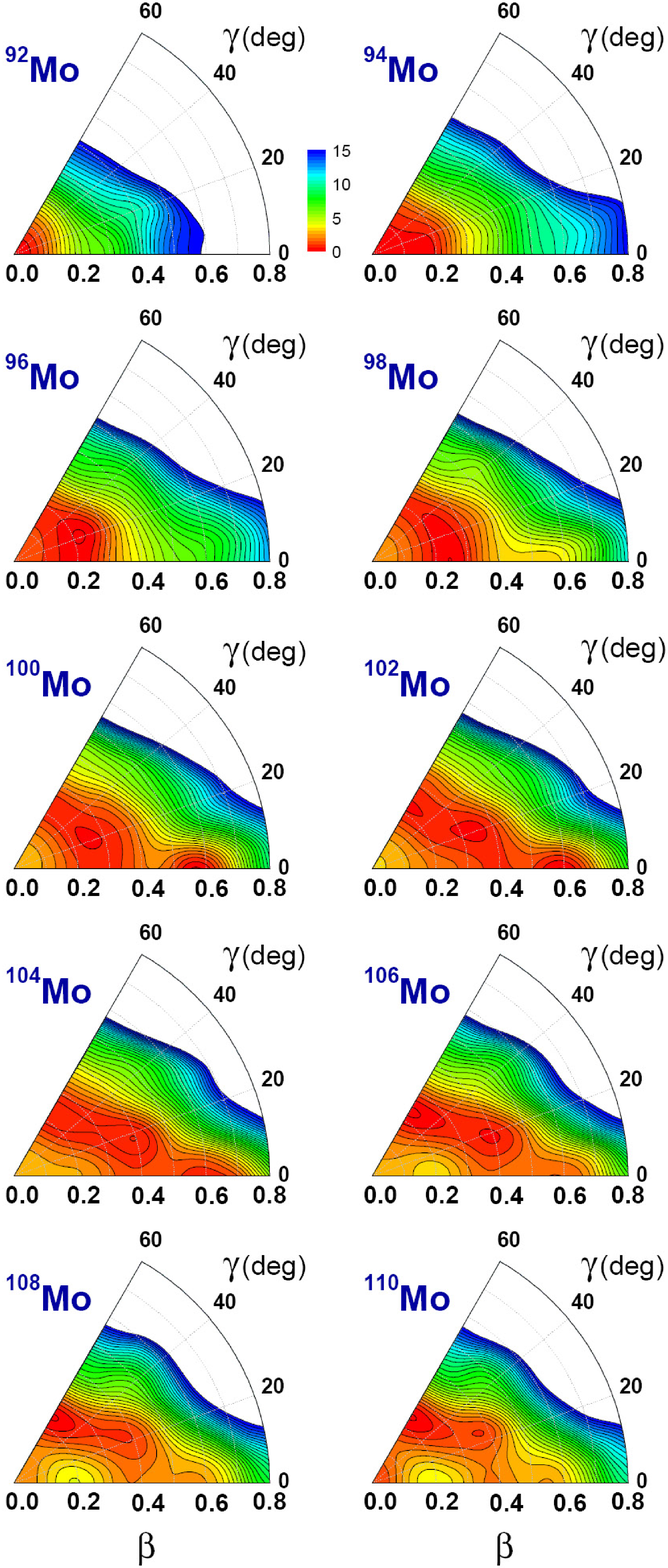}
\end{center}
\caption{\label{fig4}(Color online) Same as the Fig.~\ref{fig1}, but
for the isotopes $^{92-110}$Mo.}
\end{figure}

\subsection{Shape evolution in neutron-rich Kr, Zr, and Mo
isotopes} \label{Sec.II-2}

The PESs of neutron-rich Kr, Zr, and Mo isotopes are shown in Figs.
\ref{fig2}-\ref{fig4}, respectively. In comparison with the shape
evolution picture of Sr isotopes, the main difference is found in
the evolution of prolate minimum in Kr isotopes, where the PESs are
much softer and the prolate minima are not well developed.

The shape evolution picture of Zr isotopes is very similar as that
in Sr isotopes, except the barrier height separating the prolate and
oblate minima. In contrary with the case in Sr isotopes, the prolate
and oblate minima are always connected through triaxial distortion
with near-zero barrier height, in particular, for $^{100}$Zr with
shape coexistence phenomenon.

For Mo isotopes, the shape evolution picture is similar as that in
Zr isotopes. The evident difference is the occurrence of triaxial
minima in the Mo isotopes with neutron number from $N=58$ to $N=68$.

Very recently, a global study of nuclear low-lying states based on
the nonrelativistic Hartree-Fock-Bogoliubov framework with the Gogny
force have been done. The corresponding potential energy surfaces
and other observables are given in Ref.~\cite{CEA}. Based on the
same framework, Rodriguez-Guzm\'{a}n \emph{et al.} have examined in
detail the shapes evolution of nuclear ground-state in neutron-rich
Sr, Zr, and Mo isotopes, including both even-even and odd-A
nuclei~\cite{Guzman10}. The trend of shape evolution is similar as
our result based on the covariant density functional. However, the
transition at $N=60$ in our calculations is a little slower along
the isotopic chain, and more rapid along the isotonic chain for the
$N=60$ isotones.

\begin{figure}[]
\begin{center}
\includegraphics[width=0.40\textwidth]{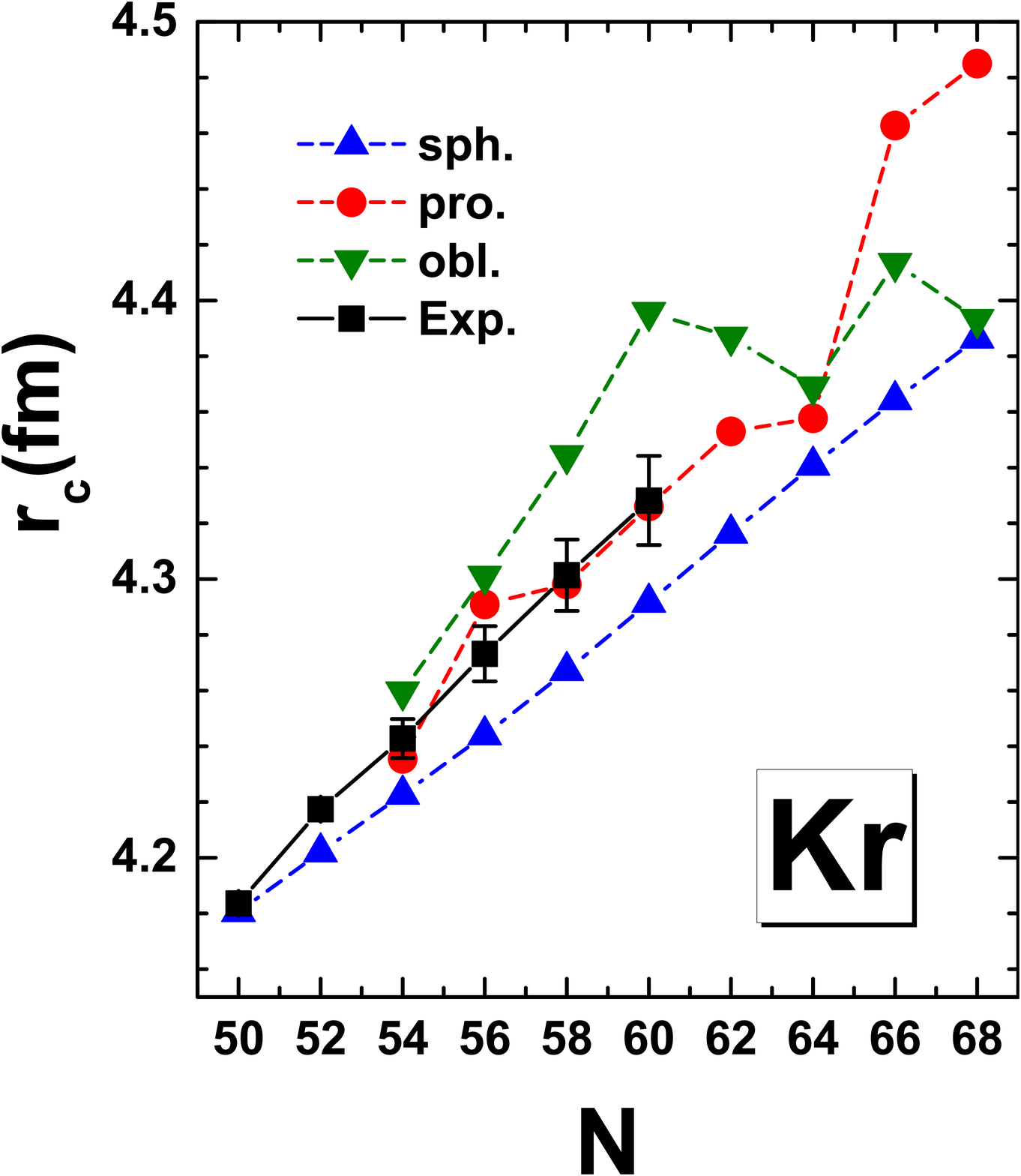}
\end{center}
\caption{\label{fig5}(Color online) Same as the Fig.\ref{fig01}, but
for Kr isotopes.}
\end{figure}

\begin{figure}[]
\begin{center}
\includegraphics[width=0.40\textwidth]{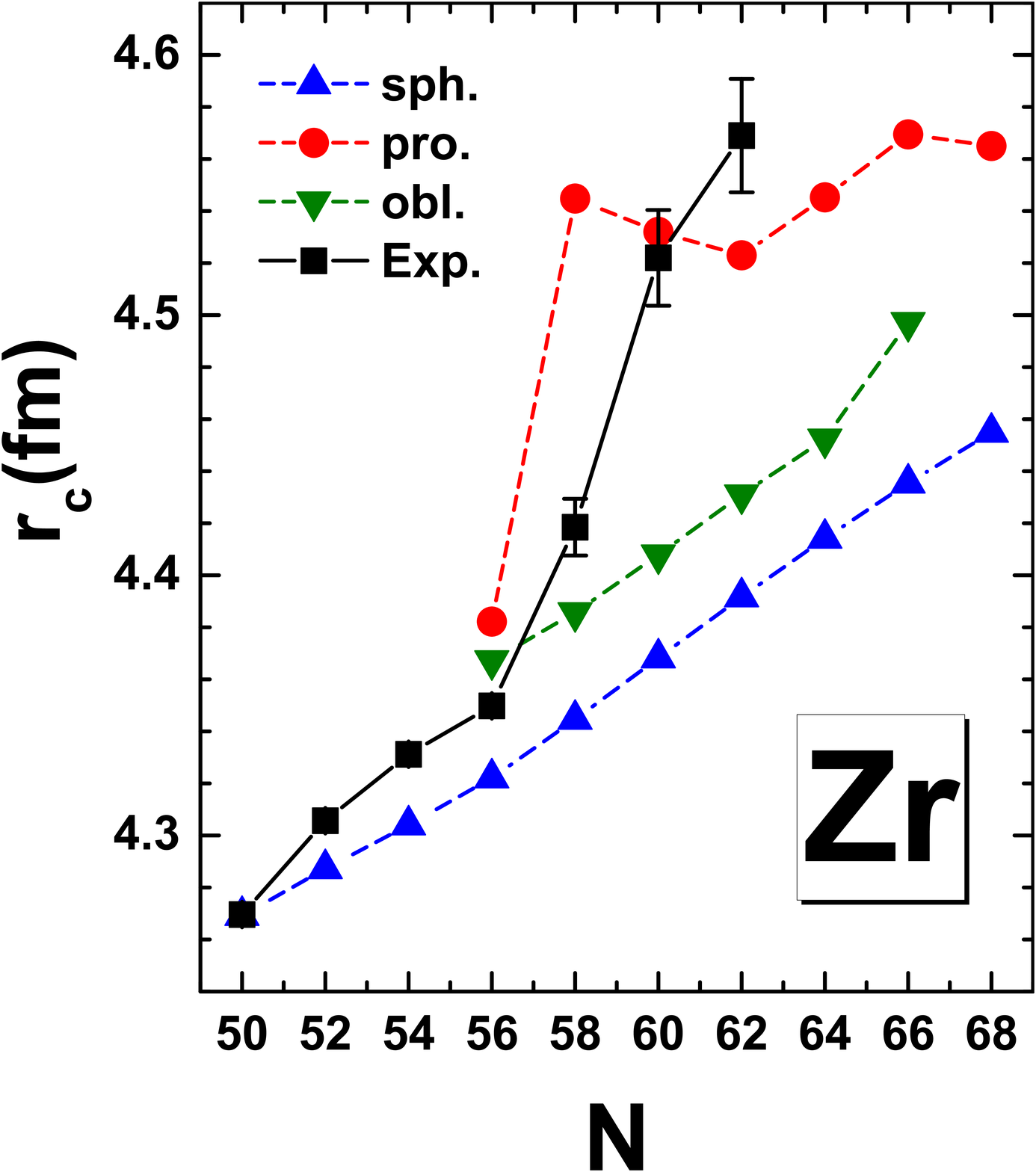}
\end{center}
\caption{\label{fig6}(Color online) Same as the Fig.\ref{fig01}, but
for Zr isotopes.}
\end{figure}

\begin{figure}[]
\begin{center}
\includegraphics[width=0.40\textwidth]{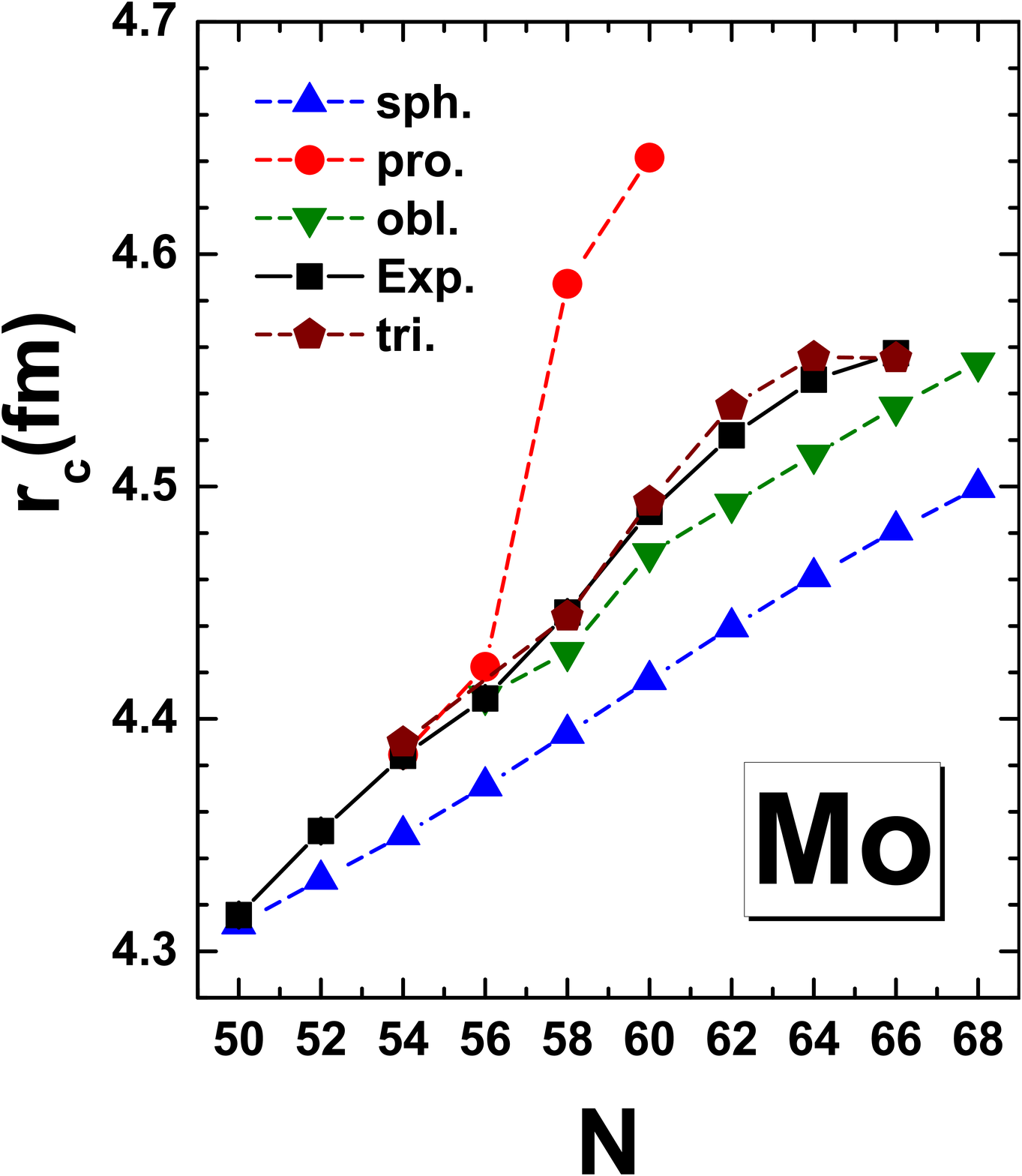}
\end{center}
\caption{\label{fig7}(Color online) Same as the Fig.\ref{fig01}, but
for Mo isotopes.}
\end{figure}

The charge radii of Kr, Zr, and Mo isotopes are plotted in
Figs.~\ref{fig5}-\ref{fig7}, respectively. The sharp transition is
also observed in Zr isotopes, which also indicates the rapid change
in the nuclear shapes. On the contrary, the charge radii in Kr and
Mo isotopes increase smoothly with the neutron number. Similar as
the calculation results of the Gogny force in Ref.~\cite{Guzman10},
the triaxiality is shown to be essential to reproduce qualitatively
the charge radii in Mo isotopes.

\subsection{Covariant density functional based 5D collective
Hamiltonian analysis of shape coexistence in $^{98}{\rm Sr}$ and
$^{100}{\rm Zr}$} \label{coexistence}

The coexistence of prolate and oblate shapes observed in $^{98}$Sr
and $^{100}$Zr will be studied in more detail with the 5D collective
Hamiltonian determined by the constrained self-consistent RMF plus
BCS calculations. The details about the covariant density functional
based 5D collective Hamiltonian can be found in
Ref.~\cite{Niksic10}.

\begin{figure}[]
\begin{center}
\includegraphics[width=0.6\textwidth]{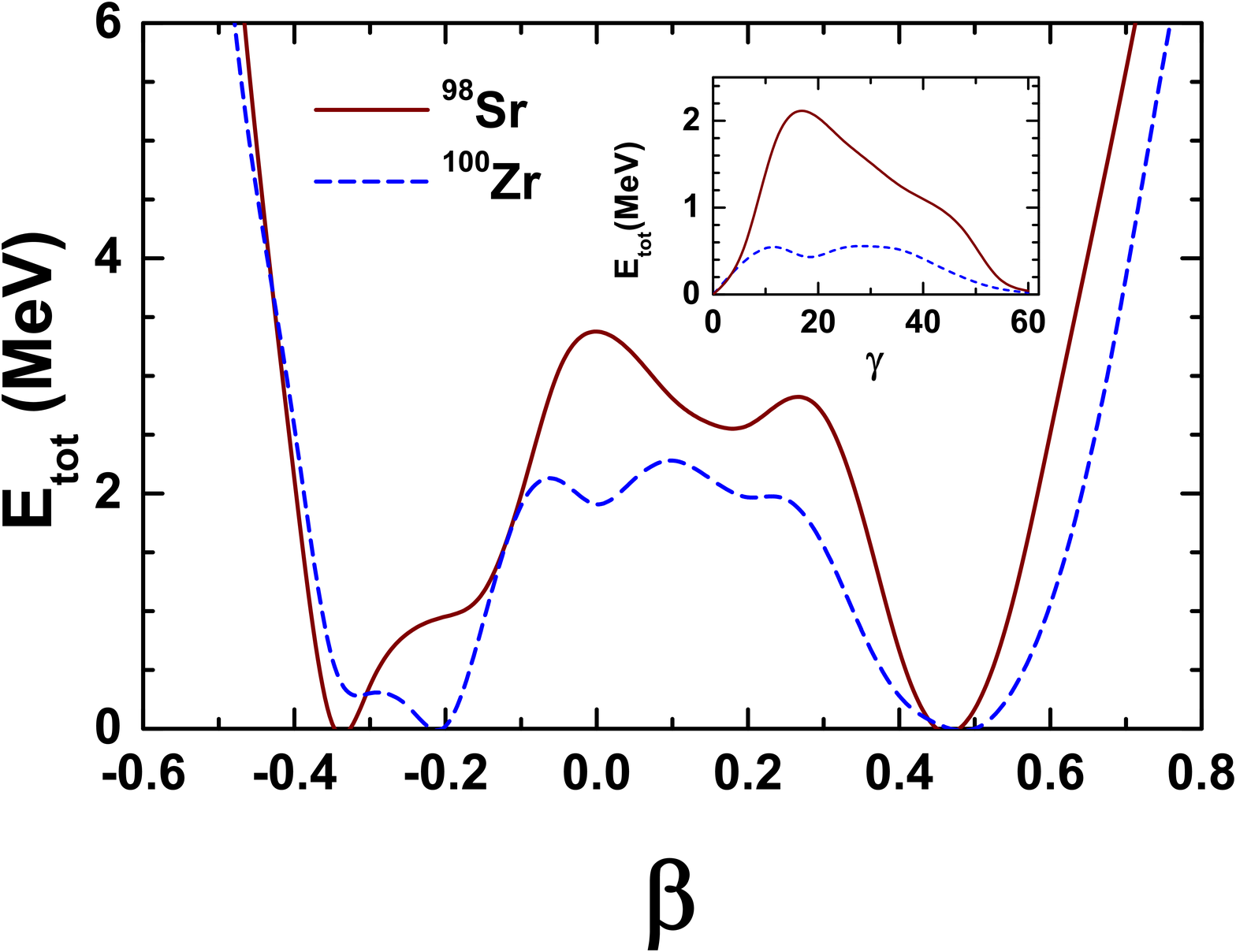}
\end{center}
\caption{\label{fig8}(Color online) The total energies of $^{98}$Sr
and $^{100}$Zr as functions of axial deformation $\beta$. All
energies are normalized with respect to the total energy of the
absolute minimum. The inset displays the PEC corresponding to the
projections on the $\gamma$ deformation, that is, the minimum for
each $\gamma$ on the PES in the $\beta$-$\gamma$ plane.}
\end{figure}

In Fig. \ref{fig8} we plot the total energies as functions of axial
deformation $\beta$ for $^{98}$Sr and $^{100}$Zr. The inset displays
the PEC corresponding to the projections on the $\gamma$
deformation, that is, the minimum for each $\gamma$ deformation on
the PES in the $\beta$-$\gamma$ plane (c.f. Figs.~\ref{fig1} and
\ref{fig3}). In both nuclei, the coexisting prolate and oblate
minima with very closed binding energies are observed, which are
separated by certain barriers. In $^{98}$Sr, the spherical barrier
height is $\sim3.4$~MeV. After considering the $\gamma$ degree of
freedom, this barrier height is lowered down to $~2.2$~MeV. In
$^{100}$Zr, the barrier height is much smaller with the size
$\sim0.5$~MeV if the $\gamma$ deformation is considered.

\begin{table}[htb!]
\centering \caption{ The calculated excitation energies (in MeV) of
$0^+_2$ states and E0 transition strengths $\rho^2(E0;
0^+_2\rightarrow 0^+_1)\times10^3$ in $^{98}$Sr and $^{100}$Zr, in
comparison with the corresponding
data~\cite{Kibedi200577,E0transition}}
\begin{tabular}{cccccc}
 \hline\hline
 \ & \multicolumn{2}{c}{\rule[-0.1cm]{0cm}{0.5cm}$^{98}$Sr} &\ & \multicolumn{2}{c}{$^{100}$Zr}\\
 \cline{2-3}\cline{5-6}
 \rule[-0.1cm]{0cm}{0.5cm}\  & Cal. & Exp. &\ & Cal. & Exp.\\
 \hline
 E($0^+_2$)(MeV)                                                     & 0.216   & 0.215 &\ & 0.468   & 0.331\\
 $\rule[-0.1cm]{0cm}{0.5cm}\rho^2(E0; 0^+_2\rightarrow 0^+_1)\times10^3$ & 116.841 & 51(5) &\ & 150.321 & 108(19)\\
\hline \hline
\end{tabular}\label{table1}
\end{table}

The excitation energy of the $0^+_2$ state and the E0 transition
strength $\rho^2(E0;{0^+_2 \to 0^+_1})$ between the $0^+_2$ and
$0^+_1$ states are two key quantities in the study of shape
coexistence,
\begin{equation}
\rho^2(E0;{0^+_2 \to 0^+_1}) =\left|\frac{\langle0^+_2|\sum_k
e_kr_k^2|0^+_1\rangle}
  {eR^2_0}\right|^2,
\end{equation}
where $R_0\simeq1.2A^{1/3}$~fm. The $\rho^2(E0;{0^+_2 \to 0^+_1})$
is related to the change in the root mean-square charge radius of
the nucleus between the $0^+_1$ and $0^+_2$ states, and therefore
carries important information about the change in deformation and
the overlap of the wave functions.

In Tab.~\ref{table1}, we list the calculated excitation energies of
$0^+_2$ states and E0 transition strengths $\rho^2(E0;
0^+_2\rightarrow 0^+_1)$ in $^{98}$Sr and $^{100}$Zr from the
solution of 5D collective Hamiltonian based on the energy functional
PC-PK1 plus the separable pairing force. The experimental
data~\cite{Kibedi200577,E0transition} are also shown for comparison.
The existence of very low-lying $0^+_2$ state is often used as a
strong signal for the shape coexistence. As expected, the calculated
excitation energies of $0^+_2$ states in both nuclei are predicted
in very low values, that is, 0.216~MeV for $^{98}$Sr and 0.468~MeV
for $^{100}$Zr, which are also very close to the data.  Although the
experimental E0 transition strengths $\rho^2(E0; 0^+_2\rightarrow
0^+_1)$ are overestimated by the collective Hamiltonian based on
PC-PK1 functional, they are all typically large, again confirming
the shape coexistence phenomena in these two $N=60$ isotones.

\begin{figure}[]
\centering
\includegraphics[scale=0.25]{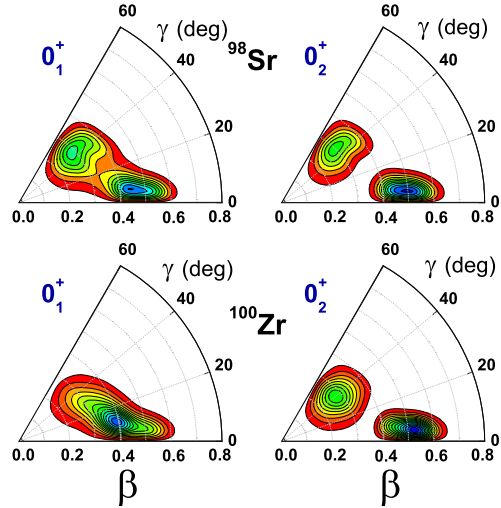}
\caption{\label{fig10}(Color online) Probability density
distribution in the $\beta$-$\gamma$ plane for the $0^+_1$ and
$0^+_2$  states of $^{98}$Sr and $^{100}$Zr.}
\end{figure}

The mixing between the $0^+_1$ and $0^+_2$ states can be further
understood from the distribution of the wave functions of the
$0^+_1$ and $0^+_2$ states. Figure~\ref{fig10} displays the
probability density distribution of $0^+_1$ and $0^+_2$ states in
$\beta$-$\gamma$ plane for $^{98}$Sr and $^{100}$Zr. Due to the
hight triaxial barrier (c.f. Fig.~\ref{fig1}), two peaks
corresponding to the coexisting prolate and oblate shapes are
observed in both $0^+_1$ and $0^+_2$ states in $^{98}$Sr. However,
in $^{100}$Zr, the probability density of the $0^+_1$ state is
almost uniformly distributed along the $\gamma$ deformation,
connecting the prolate and oblate shapes. Since there is one node in
the probability distribution of $0^+_2$ state, the resultant
$\rho^2(E0;{0^+_2 \to 0^+_1})$ is the consequence of concelation
from the probability distributions of prolate and oblate parts. This
cancelation is larger in $^{98}$Sr than that in $^{100}$Zr. As a
result, the obtained $\rho^2(E0;{0^+_2 \to 0^+_1})$ in  $^{98}$Sr is
much smaller than the value in $^{100}$Zr, as shown in
Tab.~\ref{table1}.

\begin{figure}[]
\begin{center}
\includegraphics[width=0.7\textwidth]{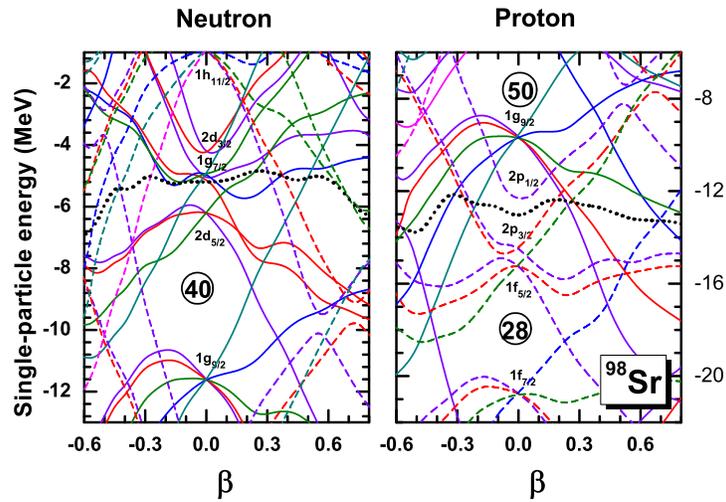}
\end{center}
\caption{\label{fig9}(Color online) Neutron and proton
single-particle levels for $^{98}$Sr as functions of the axial
deformation parameter $\beta$. The thick dotted curves denote the
corresponding Fermi levels.}
\end{figure}

The observed shape coexistence phenomenon can be understood from the
distribution of single-nucleon levels. In Fig.~\ref{fig9}, we plot
the neutron and proton single-particle energy levels in $^{98}$Sr as
functions of the axial deformation parameter $\beta$. The thick
dotted curves denote the position of the corresponding Fermi levels.
It is shown that the neutron Fermi level goes across the deformation
region of low level density with $-0.4\leq\beta\leq-0.15$, giving
rise to the oblate minimum. On the other hand, the proton Fermi
level locates in the middle of the energy gap around
$\beta\sim0.45$, which gives rise to the prolate minimum in
$^{98}$Sr.

\section{Summary}
\label{summary}
 In summary, the triaxial relativistic mean-field plus BCS model
 with a point-coupling interaction in the particle-hole channel
 and a separable pairing force in the particle-particle channel has been established
 and applied to study the shape evolution and shape coexistence phenomena in neutron-rich $A\sim100$
 nuclei, including Kr, Sr, Zr, and Mo isotopes using the newly parameterized
 PC-PK1 energy functional. The evolution of potential energy surfaces and charge
 radii with the neutron number in each isotopes have been presented.
 Sharp rising in the charge radii of Sr and Zr isotopes at $N=60$
 has been observed and shown to be related to the rapid changing in the nuclear
 shape. This dramatic evolution of charge radii is smoothed out in Mo
 isotopes due to the occurrence of triaxial minimum, which is
 similar as the results of Hartree-Fock-Bogogliubov
 calculations with the Gogny force. In particular,
 the triaxiality has been shown to be essential to
 reproduce quantitatively the charge radii of Mo isotopes.

 The coexistence of prolate and oblate shapes
 has been observed in $^{98}$Sr and $^{100}$Zr. However,
 the barrier height separating the coexisting minima
 along the $\gamma$ deformation in $^{100}$Zr has been shown
 much lower than that in $^{98}$Sr. The observed oblate minimum
 and prolate minimum are related to the low single-particle energy level
 density around the Fermi surfaces of neutron and proton
 respectively. Furthermore, the 5D collective Hamiltonian determined by the
 calculations of the PC-PK1 energy functional has been constructed
 and solved for $^{98}$Sr and $^{100}$Zr. The resultant excitation
 energy of $0^+_2$ state and E0 transition strength $\rho^2(E0;0^+_2\rightarrow0^+_1)$
 are in rather good agreement with the data. It has been found that the lower barrier
 height in $^{100}$Zr gives rise to the larger $\rho^2(E0;0^+_2\rightarrow0^+_1)$
 than that in $^{98}$Sr.

 \section*{Acknowledgments}
JMY would like to thank Peter Ring and Yuan Tian for helpful
discussions and acknowledge a postdoctoral fellowship from the
F.R.S.-FNRS (Belgium). This work was partly supported by the Major
State Basic Research Developing Program 2007 CB815000, the National
Science Foundation of China under Grants No. 10947013 and No.
10975008, the Fundamental Research Funds for the Central
Universities (XDJK2010B007), and the Southwest University Initial
Research Foundation Grant to Doctor (SWU109011 and SWU110039).



\begin{thebibliography}{99}


\bibitem{Mach91} H. Mach \emph{et al.}, Nucl. Phys. \textbf{A523} (1991) 197.

\bibitem{Goodin07}C. Goodin \emph{et al.}, Nucl. Phys. \textbf{A787} (2007) 231.

\bibitem{Urban01}W. Urban \emph{et al.},  Nucl. Phys. \textbf{A689} (2001) 605.

\bibitem{NNDC}  National Nuclear Data Center, Brookhaven National Laboratory, {\it http://www.nndc.bnl.gov/}.

 \bibitem{Hager06}U. Hager \emph{et al.}, Phys. Rev. Lett. \textbf{96} (2006) 042504.

\bibitem{Charlwood09}F. C. Charlwood \emph{et al.}, Phys. Lett. \textbf{B674}  (2009) 23.

\bibitem{Marginean09}N. Marginean \emph{et al.}, Phys. Rev. \textbf{C80} (2009) 021301(R).


 \bibitem{Keim95}M. Keim, E. Arnold, W. Borchers {\it et al}., Nucl. Phys. \textbf{A586} (1995) 219.


\bibitem{Jung80} G. Jung \emph{et al.}, Phys. Rev. \textbf{C22} (1980) 252.

\bibitem{Kawade82}K. Kawade \emph{et al.}, Z. Phys. A \textbf{304} (1982) 293.

\bibitem{Lhersonneau94}G. Lhersonneau \emph{et al.}, Phys. Rev. \textbf{C49} (1994) 1379.

\bibitem{Mach89} H. Mach \emph{et al.}, Phys. Lett.  \textbf{B230} (1989) 21.

\bibitem{Schussler80}F. Schussler, J, A. Pinston, B. Monnand and A. Moussa, Nucl. Phys.  \textbf{A339} (1980) 415.

 \bibitem{Clement10} E. Cl\'{e}ment {\it et al}., CERN-INTC-2010-009/INTC-P-216-ADD-108/01/ 2010.

\bibitem{Federman78} P. Federman and S. Pittel, Phys. Lett. \textbf{B77} (1978) 29.

\bibitem{Kumar85} A. Kumar and M. R. Gunye, Phys. Rev. \textbf{C32} (1985) 2116.

\bibitem{Galeriu86}D. Galeriu, D. Bucurescu, and M. Ivaqcu, J. Phys. G\textbf{12} (1986) 329.

\bibitem{Michiaki90}S. Michiaki and A. Akito, Nucl. Phys. \textbf{A515} (1990) 77.

\bibitem{Moller95}P. M\"{o}ler, J. R. Nix, W. D. Myers, and W. J. Swiatecki, At. Data Nucl. Data Tables \textbf{59} (1995) 185.

\bibitem{Skalski97}J. Skalski, S. Mizutori, and W. Nazarewicz, Nucl. Phys. \textbf{A617} (1997) 282.

\bibitem{Xu02}F. R. Xu, P.M. Walker, and R. Wyss, Phys. Rev. \textbf{C65} (2002) 021303(R).

\bibitem{Sonia08} S. Verma, P. Ahmad Dar, and R. Devi, Phys. Rev. \textbf{C77} (2008) 024308.

\bibitem{Gracia05}J. Garc\'{i}a-Ramos, K. Heyde, R. Fossion, V. Hellemans, and S. De Baerdemacker,
                                     Eur. Phys. J. \textbf{A26}  (2005) 221.

\bibitem{Sieja09} K. Sieja, F. Nowacki, K. Langanke, and G. Mart\'\i{}nez-Pinedo, Phys. Rev. \textbf{C79} (2009) 064310.


\bibitem{Bonche85}P. Bonche, H. Flocard, P. H. Heenen, S. J. Krieger, M. S. Weiss, Nucl. Phys. \textbf{A443} (1985) 39.

\bibitem{Skalski93}J. Skalski, P.-H. Heenen, and P. Bonche, Nucl. Phys. \textbf{A559} (1993) 221.

\bibitem{Bender06} M. Bender, G. F. Bertsch, P.-H. Heenen, Phys. Rev. {\textbf C73} (2006) 034322; Phys. Rev. {\textbf C78} (2008) 054312.

\bibitem{Delaroche10} J. -P. Delaroche, M. Girod, J. Libert {\it et al.,} Phys. Rev. \textbf{C81} 014303 (2010).
\bibitem{CEA} S. Hilaire and M. Girod, {\it http://www-phynu.cea.fr/ science\_en\_ligne/
              carte\_potentiels\_microscopiques/carte\_potentiel\_nucleaire.htm.}

\bibitem{Guzman10}R. Rodr\'{i}guez-Guzm\'{a}n, P. Sarriguren, L. M. Robledo and S. Perez-Martin,
                                         Phys. Lett. \textbf{B691} (2010) 202.

\bibitem{Lalazissis99} G. A. Lalazissis, S. Raman, P. Ring, At. Data Nucl. Data Tables \textbf{71} (1999) 1.

 \bibitem{Reinhard89}    P. G. Reinhard, Rep. Prog. Phys. {\bf52} (1989) 439.

 \bibitem{Ring96}        P. Ring, Prog. Part. Nucl. Phys. {\bf37} (1996) 193.

 \bibitem{Vretenar05}    D. Vretenar, A.~V. Afanasjev, G.~A. Lalazissis, and P. Ring, Phys. Rep. {\bf 409}  (2005)  101.

  \bibitem{Meng06}       J. Meng,  H. Toki, S.-G. Zhou, S.-Q. Zhang, W.-H. Long, and L.-S. Geng, Prog. Part. Nucl. Phys. {\bf 57}  (2006) 470.

 \bibitem{Niksic11}      T. Nik\v{s}i\'{c}, D. Vretenar and P. Ring, Prog. Part. Nucl. Phys. {\bf 66} (2011) 519.


\bibitem{Yao09}J. M. Yao, J. Meng, P. Ring,  and D. Pena Arteaga, Phys. Rev. \textbf{C79} (2009) 044312.

\bibitem{Niksic06} T.~Nik\v{s}i\'{c}, D.~Vretenar, and P.~Ring, Phys. Rev. {\textbf C74} (2006) 064309.

\bibitem{Yao10} J. M. Yao, J. Meng, P. Ring, and D. Vretenar, Phys. Rev. \textbf{C81} (2010) 044311.


\bibitem{Yao11} J. M. Yao, H. Mei, H. Chen, J. Meng, P. Ring, and D. Vretenar, Phys. Rev. \textbf{C83}  (2011) 014308.


 \bibitem{Niksic09} T. Niksic, Z. P. Li, D. Vretenar, L. Prochniak, J. Meng, and P. Ring, Phys. Rev. \textbf{C79} (2009) 034303.

\bibitem{Burvenich} T. B\"{u}rvenich, D. G. Madland, J. A. Maruhn, and P.-G. Reinhard,
                     Phys. Rev. \textbf{C65}, 044308 (2002).

\bibitem{Niksic}T. Nik\v{s}i\'{c}, D. Vretenar, and P. Ring, Phys. Rev. \textbf{C78}  (2008) 034318.

\bibitem{Zhao10} P. W. Zhao, Z. P. Li, J. M. Yao, and J. Meng, , Phys. Rev. \textbf{C82}  (2010) 054319.

\bibitem{Tian091} Y. Tian and Z. Y. Ma, and P. Ring, Phys. Lett. \textbf{B676} (2009) 44.

\bibitem{Tian092} Y. Tian, and Z. Y. Ma,  and P. Ring, Phys. Rev. \textbf{C79}  (2009) 064301.

\bibitem{Tian093} Y. Tian, and Z. Y. Ma, and P. Ring, Phys. Rev. \textbf{C80}  (2009) 024313.

\bibitem{Niksic10}T. Nik\v{s}i\'{c}, P. Ring, D. Vretenar, Y. Tian, and Z. Y. Ma, Phys. Rev. \textbf{C81}  (2010) 054318.

 \bibitem{Li10}  Z. P. Li, T. Nik\v{s}i\'{c}, D. Vretenar, P. Ring, and J. Meng, Phys. Rev. \textbf{C81}  (2010) 064321.

\bibitem{Ring80} P. Ring and P. Schuck, {\em The Nuclear Many-Body Problem}
(Springer, Heidelberg, 1980).
\bibitem{Angeli} I. Angeli, At. Data Nucl. Data Tables \textbf{87}  (2004) 185.

\bibitem{Kibedi200577}T. Kib\'{e}di and R. H. Spear, At. Data Nucl. Data Tables \textbf{89}  (2005) 77.

\bibitem{E0transition} {\it http://ie.lbl.gov/TOI2003/GammaSearch.asp.}


\end{thebibliography}
\end{document}